# The Development Strategy of IT Capability: A Contingency Perspective

**Po-Yen Chen***
Department of Information Management
National Taiwan University
Taipei, Taiwan
Email: poyen.taiwan@gmail.com

**Chorng-Shyong Ong**
Department of Information Management
National Taiwan University
Taipei, Taiwan
Email: ongcs@ntu.edu.tw

## Abstract

This study proposes a conceptual model to link IT capabilities, industry types, and value implications. We attempt to use a contingency analysis to theorize that which types of IT capabilities (e.g., externally-focused, internally-focused, and aggregate IT capability) should a firm develop and then what benefits (e.g., firm value and firm performance) it will gain according to its industry's value creation logic (e.g., value chain-based, value shop-based, and value network-based industry). The empirical findings show that a value network-based firm should develop externally-focused IT capabilities to create its firm value and a value chain-based firm should develop aggregate IT capabilities to improve its firm performance and create its firm value.

## 1 Introduction

Understanding the business value of information technology (IT) has been a critical issue to IS researchers and practitioners and has been one of the major research topics in IS field (Kohli and Grover 2008; Schryen 2013). In the early stage, a rich body of literature has dedicate to examine the relationship between IT spending and productivity (or other economic impact) but the empirical results have reported mixed findings (Dedrick et al. 2003; Lu and Jinghua 2012; Melville et al. 2004). The phenomenon are so-called "productivity paradox" (Brynjolfsson and Hitt 1998). Drawing on the resource-based view (RBV), subsequent IS researchers have argued that superior IT capability (unlike high IT spending) can render a firm a significant advantage over its competitors (Bharadwaj 2000; Bhatt and Grover 2005; Ravichandran and Lertwongsatien 2005; Santhanam and Hartono 2003). Since Bharadwaj (2000) proposed the concept of IT capabilities, IT capabilities seem to be regarded as a general development strategy of IT for all firms and many studies have also confirmed that superior IT capability indeed can enhance a firm's performance (Bharadwaj 2000; Santhanam and Hartono 2003).

However, this argument is being challenged and re-thought again by recent studies (e.g., Chae et al. 2014; Ong and Chen 2014; Schryen 2013). Among them, Chae et al. (2014) did not find discernable evidence for the relationship between IT capability and firm performance after 2000 and Schryen (2013) notes that the value creation process remains unclear in term of how, why, and when IS assets and organizational capabilities are transformed into competitive strength. Although Chae et al. (2014) propose some possible reasons (e.g., highly homogeneous IT), they also agree that the findings should not imply that the relationship no longer exists and a firm does not need to develop its IT. Given the consensus of the importance of IT, we believe that a critical issue for practitioners is what things can a firm do when IT capability doesn't always matter.

Drawing on the work of Schryen (2013) and Wade and Hulland (2004), a contingency perspective may be advanced to address this issue. It can help us further understand the appropriate conditions of information technology (Dehning et al. 2003) and the value implication of investments in IT (Anderson et al. 2006). Similar, Muhanna and Stoel (2003) also noted that the impact of IT capabilities on firm performance was contingent on the "fit" between the types of IT capability a firm possessed and the industry environment in which it competes. Unfortunately, previous studies usually tell us fragmented knowledge and it is hard to be concluded as a consolidated development strategy of IT capability for various industries. This study will develop a conceptual model on prior literature and test the same. We first link the concepts of IT capabilities, industry types, and value implications. Then, we attempt to use a contingency perspective to test that which types of IT capabilities [i.e., aggregate IT capabilities, externally-focused IT capabilities, or internally-focused IT capabilities (Stoel and Muhanna 2009)] should a firm develop and then what benefits [i.e., firm performance or firm value (Kohli et al. 2012; Ong and Chen 2014)] it will gain according to its industry's value creation logic [i.e., value chain-based, value network-based, or value shop-based industries (Stabell and Fjeldstad 1998)].

Overall, this study has three objectives. First, we use a value implication viewpoint to illustrate various IT capabilities may bring different outcomes for firms. Previous studies usually use an overall viewpoint to measure the effects of IT and are hard to link types of IT capabilities to their benefit effects on firms. Second, industry roles are included to review the fitness between industry's value creation logic and types of IT capabilities. From the viewpoint of resource allocation, we believe that various industries should build their unique portfolio of IT capabilities to maximize their IT value. Third, we attempt to answer our final research question, i.e., what benefits will a firm gain if it follows the best development strategy of IT capabilities. This new conceptual model can help a firm to build a more specific development strategy of IT capability.

## 2. Literature Review and Hypotheses Development

More and more studies believe that IT value is contingent on some conditions and introduce the viewpoint of strategic fit to IT value field (e.g., Stoel and Muhanna 2009) and they concern the question, "Under what conditions does IT pay off?", rather than "Does IT pay off?" (Schryen 2013). Firms are expected to achieve better performance with environmental and internal consistency, or fit, among strategic, structural, and contextual variables (Wanger et al. 2012). In this section, some important fit factors (e.g., IT capability types, value implications, and industries) are further clarified to address the contingent perspective. From the contingent perspective, we believe that IT capability may not always matter and it should match some conditions. In doing so, maybe some specific development strategy of IT capability can be derived.

### 2.1 IT Capabilities and their Classification

In general, IT capabilities can be defined as complex bundles of IT-related resources, skills and knowledge, exercised through business processes, that enable firms to coordinate activities and make use of the IT assets to provide desired results (Bharadwaj 2000; Stoel and Muhanna 2009). These aggregate IT capabilities are believed to have positive effects on firms' economic outcomes, including short-term operating benefits (i.e., firm performance) (Bharadwaj 2000; Ong and Chen 2014; Santhanam and Hartono 2003) and long-term intangible and growth value (i.e., firm value) (Bharadwaj et al. 1999; Kohli et al. 2012; Ong and Chen 2014). Thus, H1a is proposed.

$H_{1a}$: Higher **aggregate** IT capabilities will have a positive effect on **firm performance**.

$H_{1b}$: Higher **aggregate** IT capabilities will have a positive effect on **firm value**.

Further, IT capabilities can be further classified as various types (e.g., Bhatt and Grover 2005; Dehning and Stratopoulos 2003; Ravichandran and Lertwongsatien 2005; Stoel and Muhanna 2009; Wade and Hulland 2004). Among them, Stole and Muhanna (2009) further distinguished between externally and internally focused IT capabilities based on the primary business process area supported, thus reflecting the firm's choices of where and how IT resources were deployed. This shows that firms should develop their unique portfolio of IT capabilities according to the concept of fit or alignment. Externally-focused IT capabilities are defined as bundles of IT-related resources, skills and knowledge that help the firm sense and respond in a timely way to changes in its markets and shifts in the needs of customers and suppliers (Stole and Muhanna 2009). This type of capability is similar to Wade and Hulland (2004)'s outside-in IT resources, which include external relationship management and market responsiveness. Internally-focused IT capabilities are defined as bundles of IT-related resources, skills and accumulated knowledge that help the firm offer reliable products and services and minimize overhead costs (back-office production, operational support, and fulfillment processes) (Stole and Muhanna 2009). This type of capability is similar to Wade and Hulland (2004)'s inside-out IT resources, which include IS infrastructure, IS technical skills, IS development, and cost efficient IS operations. From a high level perspective, various types of capabilities may all have positive benefits (e.g., firm performance and firm value) for firms even if a firm develops non-fit IT capabilities or uses non-fit indicators to measure them. Thus, we posit that:

$H_{1b}$: Higher **externally-focused** IT capabilities will have a positive effect on **firm performance**.

$H_{1b}$: Higher **externally-focused** IT capabilities will have a positive effect on **firm value**.

    $H_{1c}$: Higher **_internally-focused_** IT capabilities will have a positive effect on **_firm performance_**.

    $H_{1c}$: Higher **_internally-focused_** IT capabilities will have a positive effect on **_firm value_**.

## 2.2 IT Capabilities and Value Implications

We should further think the roles of various IT capabilities and their value implications. As Bhatt and Grover (2005) noted that the quality of the IT infrastructure (one of IT capability types) is not related to a firm's advantage, we believe that not all types of IT capabilities have the same effects on firm performance or other indicators. Indeed, how to capture the benefits of IT is a key issue. In recent years, the important implications of dependent variables (i.e., a firm's economic impact) have been identified (e.g., Kohli et al. 2012; Mithas et al. 2012; Ong and Chen 2014). Kohli et al. (2012) found out that IT-enabled accounting measures is not significant but IT-enabled market value is significant. This shows that the argument about "IT does matter" or "IT doesn't matter" is highly related to the measurement tools. According to the methods of measurements, accounting-based measures and financial market-based measures are two main measures that have been usually used to capture IT value (Bharadwaj et al. 1999). Further, their implications are also discussed by many IT value literatures (e.g. Kohli et al. 2012; Ong and Chen 2014; Saeed et al. 2005; Tanriverdi 2006). Overall, accounting-based measures are historical (Bharadwaj et al. 1999), backward-looking measures (Tanriverdi 2006), and retrospective (Kohli et al. 2012) measures and can be referred to "firm performance" (Ong and Chen 2014). Financial market-based measures are future (Bharadwaj et al. 1999), forward-looking (Tanriverdi 2006), and prospective (Kohli et al. 2012) measures and can be referred to "firm value" (Ong and Chen 2014).

Next, if the implications of firm performance and firm value are further linked to externally-focused and internally-focused IT capabilities, we can find out that externally-focused IT capabilities are more relevant to firm value and internally-focused IT capabilities are more relevant to firm performance. As the definition of the externally-focused IT capabilities, they can help firms to quickly respond market and build their market positions (Stole and Muhanna 2009). However, these benefits are more intangible and strategic. They influence a firm's long-term development and need more time to be realized their value. Thus, the externally-focused IT capabilities may be hard to significantly reflect on short-term operating outcomes and we can infer that externally-focused IT capabilities are more relevant to IT-enabled firm value, which implies a firm's strategic positioning (Tallon et al. 2000) and how IT support for business strategy (Rivard et al. 2006). We posit that:

    $H_{2a}$: Higher **_externally-focused_** IT capabilities will have a more positive effect on **_firm value_** than on firm performance.

On the other hand, as the definition of the internally-focused IT capabilities, they intend to integrate a firm's internal operations and data and thus increase efficiency and reliability (Stole and Muhanna 2009). These benefits can be controlled by a firm itself. They have low uncertainty and may be realized quickly. Relatively, these process improvement may be hard to significantly influence a firm's long-term market value because they lake strategic or structure changes. Thus, we can infer that internally-focused IT capabilities are more relevant to IT-enabled firm performance, which implies how IT support for firm assets (Rivard et al. 2006) and enhance operational effectiveness (Tallon et al. 2000). We posit that:

    $H_{2b}$: Higher **_internally-focused_** IT capabilities will have a more positive effect on **_firm performance_** than on firm value.

## 2.3 Industry Roles in IT Value

Industry-level factors are increasingly important for IS activities (Chiasson and Davidson 2005). Some studies have shown interest in investigating the influence of IT in different industries. For example, Dos Santos et al. (1993) and Im et al. (2001) used the information intensive view to contend that IT investment has a more significantly positive impact on firm value of information-intensive industries; Anderson et al. (2006) and Dehning et al. (2003) adopted the IT strategy role view to further establish the relationship between IT investment and the IT strategy role of various industries; Mithas et al. (2013), Stole and Muhanna (2009) and Xue et al. (2012) have focused on the relationship between IT and industries' external environments. These studies usually draw on the viewpoint of IT-related theory to explain the appropriate conditions of IT and seldom differentiated that individual industry may have its own value creation logic. Indeed, at the industry level, the results are less clear (Schryen 2013) and lack a more complete framework to develop corresponding IT strategy for various industries. As Piccoli and Ives (2005) believe that the concept of competitive advantage is rooted in the logic of value creation and distribution, we attempt to use Stabell and Fjeldstad's (1998) three value creation logics (i.e., value chain, value shop, and value network) as a new classification methodology of industry roles. These three value creation logics have their own work modes and characteristics, and have been considered to be highly influential in the field of business and management (Laffey and Gandy 2009).

According to Stabell and Fjeldstad's (1998) definitions, first, the value chain is a sequential value-added from the input to the output of products. The manufacturing industry is a typical example. Their systems must to both address the internal production processes and the linkages with their suppliers and customers. Second, the value creation logic of the value shop is meant to solve customers' problems. Professional services (e.g., healthcare and consultant industries) are typical examples. They use their own knowledge to solve customers' problems. Therefore, we can infer that this type of industries focus more on the internal processes of knowledge rather than on the external linkages. Third, the value network is the value creation model of the platform. It serves as a mediator for customer linkage and must deal with the two sides of customers simultaneously (e.g., financial industries and auction industries). The length of these industries' internal processes are usually less than value chain-based industries' and value shop-based industries'. However, they pay more attentions on how to attract a large number of customers to their platform because the scale of a firm's customers may be the key success factor for this type of industries.

## 2.4 The Development Strategy of IT Capabilities

Based on the definition of three value creation logics, their best choice of development of IT capabilities can be linked and concluded. We believe various value creation logics-based industries may need different types of IT capabilities and attempt to classify them. First, the key issue of firms in the value network-based industries may be how to attract more customers and how to maintain the relationships with them (Stabell and Fjeldstad 1998). As the definition of the externally-focused IT capabilities, they focus on a firm' external processes and can help a firm to respond its customers' need. Therefore, the externally-focused IT capabilities and the value network-based industries can be matched. Further, if a firm dedicate to develop its externally-focused IT capabilities, it will gain more IT enabled-firm value according to the argument of H2a. That is, the externally-focused IT capabilities can enable some growth potential and help a firm attract more customers for the value network-based firms. These potential linkages of customers may not directly reflect on current operating profits, but they may reflect on firms' future or intangible value. Thus, the first development strategy of IT capabilities is proposed for the value network-based industries and we

$H_{3a}$: *The best development strategy for firms that are classified as the **value network-based** industries is to develop higher **externally-focused** IT capabilities and gain superior **firm value**.*

Second, as for the value shop-based industries, their core business and core value is how to use their professional knowledge to solve customers' problems. These industries focus on their internal processes and are problem-solving oriented instead of value network-based industries' market-accessing oriented. The internally-focused IT capabilities and the value shop-based industries can be matched. According to the definition of the internally-focused IT capabilities, they can help these firm to process their problem more efficiency and avoid redundant operating costs. Further, if a firm dedicate to develop its internally-focused IT capabilities, these capabilities can improve its internal processes and provide higher qualified solutions for customers and then gain IT-enable firm performance (e.g., higher revenue or lower cost) according to the argument of H2b. Thus, we posit the counter development strategy of IT capabilities for the value shop-based industries.

$H_{3b}$: *The best development strategy for firms that are classified as the **value shop-based** industries is to develop higher **internally-focused** IT capabilities and gain superior **firm performance**.*

Third, the value chain-based firms may be the most complex structure in the three value creation logics. They must to operate internal processes (e.g., production systems) and maintain external relationships (e.g., customers and suppliers) at the same time. We believe that this type of industries must to balance their IT development and may need to develop these two types of IT capabilities at the same time. That is, they are encouraged to develop higher aggregate IT capabilities, which include higher externally-focused and internally-focused IT capabilities. And then, they can gain more comprehensive benefits, which include superior firm performance and firm value. Thus, the last development strategy is proposed for the value chain-based industries.

$H_{3c}$: *The best development strategy for firms that are classified as the **value chain-based** industries is to develop higher **aggregate** IT capabilities and gain superior **firm performance** and **firm value**.*

According to our hypothesis development, the conceptual model is further proposed as Figure 1 to illustrate the relationship between IT capabilities, industries, and firms' outcomes.

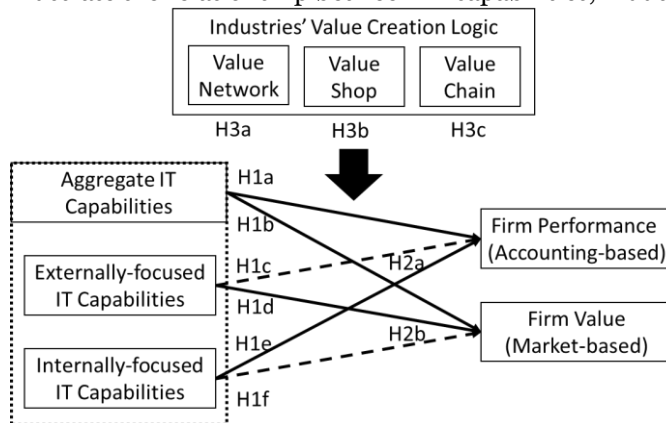

*Figure 1. Conceptual Model*

# 3 Methodology

## 3.1 The Selection of IT Leader Sample

To identify firms with superior IT capability (i.e., IT leaders), the annual list of the InformationWeek 500 (IW 500) is used. InformationWeek is an important magazine that surveys the use of IT among U.S. companies (Bharadwaj et al. 1999). In recent years, many studies identify firms with high IT capability, using the IW 500 ranking (e.g., Bharadwaj 2003; Chae et al. 2014; Lim et al. 2012; 2013; Mishra et al. 2013; Muhanna and Stoel 2010; Santhanam and Hartono 2003). IW 500 rankings benchmarks a firm's technology strategies and practices against leading organizations operating in the U.S. (Mishra et al. 2013). To creating the rankings, IW identifies the firms that are the most innovative users of business technology. Given that largest and most innovative IT organizations must demonstrate a pattern of technological, procedural, and organizational innovation to be included in the IW lists, IW rankings have been used as a reliable proxy for firms' IT capability (Bharadwaj 2003; Chae et al. 2014; Santhanam and Hartono 2003).

During 2000 to 2001, IW 500 provided an extra survey that ranked each of IW 500 IT leaders based on its creative use of technologies that can help create significant business value in two categories: e-business and technology enabled business practices. Although IW500 stopped this survey after 2002, it provides us a valuable opportunity to gather various IT capability data to conduct a large scale and objective analysis. InformationWeek described the e-business technology innovation category as focusing on e-commerce and supply-chain management. Both involved the deployment of innovative externally facing systems to serve customers and suppliers. Therefore, this category can be well represented as a proxy for the firm's externally-focused IT capabilities (Stoel and Muhanna 2009). On the other hand, Information Week described the technology enabled business practices category as focusing on ERP implementation and IS integration across business processes. Such efforts were believed to typically aim at improving the ability to manage the day-to-day operations and enhance overall efficiency and were regarded as a proxy for the firm's internally-focused IT capabilities (Stoel and Muhanna 2009).

IW only ranked the firms' innovative use of IT in e-business and business practices by assigning them a gold, silver or bronze medal in each of these categories and the rankings were made on a curve and, for each category, a third of the companies received gold, a third received silver, and the others received bronze medals. For a more conservative reason, only firms that are assigned gold medals in each of these categories in 2000 and 2001 are selected as the leader samples. To the consistent standard, we also only select the former one third of the IW 500 (i.e., rank 1 to rank 166) as the leader sample of the aggregate IT capabilities.

Next, because firms' financial data should be collected, only firms that are included in the Compustat financial database are retained. Some firms cannot be identified because they have not entered the market, their names have been changed or they are not included in the database, and so on. In addition, because the IW lists are published every September, the previous year's financial data are used in our analysis.

On the other hand, the selection procedures of the control samples also should be done. Santhanam and Hartono's (2003) matching procedures are followed. First, firms in the Compustat database that have never been recognized on IW 500 lists from 2000 to 2001 are regarded as the potential firms of control lists (i.e., they are regard as having no any superior IT capabilities that include aggregate, externally-focused, and internally-focused IT capabilities). Second, to be consistent with the leader samples, only the firms whose revenues are over 1,000 million dollars can be on the control group list. This helps avoid any bias related to firm size. Finally, because the NAICS system is defined according to a production-oriented concept and because firms in the same industry use closely similar technology, the matched industrial sample method is very appropriate for the selection of the control sample. This study follows Santhanam and Hartono's (2003) method for calculating the industry average (we use median rather than mean in order to avoid the impact of extreme numbers) according to 3-digit-NAICS (99 sub-sectors). And then, the leader samples and control samples are one-to-one matched according to industry sub-sectors. The leader samples and the control samples (after matching) are further used as testing samples in the following section. In sum, there are 180 pair-firms in the group of the aggregate IT capabilities, 185 pair-firms in the group of the externally-focused IT capabilities, and 163 pair-firms in the group of the internally-focused IT capabilities.

## 3.2 The Classification of Industries

According to Stabell and Fjeldstad (1998) and Gottschalk's (2006) classifications, traditional manufacturing (NAICS: 310000-339999), oil and gas extraction (NAICS: 210000-213999), and food chain (NAICS: 720000-729999) are in the configuration of the value chain; law, architecture, hospitals, education, and other professional service (NAICS: 511000-511999; 518000-518999; 540000-629999) are in the configuration of the value shop; communication companies (NAICS: 517000-517999), financial service (NAICS: 520000-539999), public utilities (NAICS: 220000-229999), transportation (NAICS: 480000-499999), and distribution companies (NAICS: 420000-459999) are in the configuration of the value network. After the classification procedure, the sample sizes of each value creation logics are reported in Table 2 to Table 4.

## 3.3 The Measurement of Firm Performance and Firm Value

Return on assets (ROA), which has been the most common indicator of the accounting-based measures in previous studies (e.g., Dehning and Richardson 2002), is adopted to represent firm performance in this study. As for the market-based measures, the market value to book value ratio (Tobin's q; Q ratio) has also been used by many IT value literature (e.g., Bharadwaj et al. 1999; Kohli et al. 2012). It is usually regard as a growth indicator for future book value and can be used as the measurement indicator of firm value. The Q ratio's calculation method can refer to Bharadwaj et al.'s (1999) method. That is, Q ratio = (market value + the value of preferred stock + DEBT) / total book assets. This ratio is also adopted as the proxy for firm value in this study.

# 4 Results

Based on Santhanam and Hartono (2003) and Chae et al.'s (2014) regression analysis, prior performance or value (ROA(t-1) or Q(t-1)) is used as control variables to adjust prior financial performance to avoid financial halo effects. Data year should also be controlled (Stoel and Muhanna 2009) and one dummy variable (Dummy(year)) is used. Next, in order to test the firm performance's and firm value's differences between leader groups and control groups, another dummy variable (Dummy(group)) is used to distinguish leader groups (Dummy(group)=1) and control groups (Dummy(group)=0). The results are shown in Table 1 according to each types of IT capability.

For all firms, overall, we can find out that the estimate of Dummy (group) of ROA and Q ratio are both positive in the scenario of the aggregate IT capabilities and their p-value are also significant (ROA: 0.100; Q ratio: 0.067). This shows that the aggregate IT capabilities have significantly positive effects on firm performance and firm value and then the H1a and H1b are supported. Similar, as for the scenario of the externally-focused IT capabilities, H1c and H1d are also supported (ROA: 0.076; Q ratio: 0.022). Unfortunately, as for the scenario of the internally-focused IT capabilities, H1e and H1f are not supported (ROA: 0.485; Q ratio: 0.244).

Further, the significant level of different dependent variables will be compared (i.e., H2a and H2b). Although both ROA and Q ratio are significant in the scenario of the externally-focused IT capabilities, the significant level of Q ratio (0.022) is higher than that of ROA (0.076). That is, the externally-focused IT capabilities-enabled firm value is more significant than the externally-focused IT capabilities-enabled firm performance. H2a is supported. However, because both ROA and Q ratio are not significant in the scenario of the internally-focused IT capabilities, they cannot be further compared and H2b is not supported.

| Types of IT  Parameter | Aggregate IT Capabilities | | Externally-focused IT Capabilities | | Internally-focused IT Capabilities | |
|---|---|---|---|---|---|---|
| | ROA | Q Ratio | ROA | Q Ratio | ROA | Q Ratio |
| Intercept | 0.034 (<.001***) | 0.364 (<.001***) | 0.026 (<.001***) | 0.321 (<.001***) | 0.035 (<.001***) | 0.058 (0.540) |
| ROA$_{(t-1)}$ or Q$_{(t-1)}$ | 0.373 (<.001***) | 0.789 (<.001***) | 0.371 (<.001***) | 0.765 (<.001***) | 0.353 (<.001***) | 1.026 (<.001***) |
| Dummy$_{(year)}$ | -0.13 (0.034**) | -0.290 (0.005***) | -0.003 (0.727) | -0.227 (0.021**) | -0.016 (0.016**) | -0.294 (0.004***) |
| **Dummy$_{(group)}$** | **0.010 (0.100*)** | **0.197 (0.067*)** | **0.014 (0.076*)** | **0.233 (0.022**)** | **0.005 (0.485)** | **0.120 (0.244)** |
| F statistics | 46.54 (<.001***) | 281.14 (<.001***) | 20.66 (<.001***) | 291.84 (<.001***) | 23.93 (<.001***) | 342.75 (<.001***) |
| R-sq. | 0.282 | 0.703 | 0.145 | 0.705 | 0.182 | 0.762 |
| Adjusted R-sq. | 0.276 | 0.701 | 0.138 | 0.703 | 0.175 | 0.760 |
| N | Leader:180; Control:180 | | Leader:185; Control:185 | | Leader:163; Control:163 | |

*Note: Significance at: * 10%; ** 5%; *** 1%; Dummy$_{(group)}$: leader firm=1, control firm=0; Dummy$_{(year)}$: year 2000=0, year 2001=1.*

*Table 1. The Results of All Firms*

| Types of IT  Parameter | Aggregate IT Capabilities | | Externally-focused IT Capabilities | | Internally-focused IT Capabilities | |
|---|---|---|---|---|---|---|
| | ROA | Q Ratio | ROA | Q Ratio | ROA | Q Ratio |
| Intercept | 0.012 (0.061*) | 0.538 (<.001***) | 0.023 (0.002***) | 0.461 (<.001***) | 0.015 (0.066*) | 0.518 (0.001***) |
| ROA$_{(t-1)}$ or Q$_{(t-1)}$ | 0.859 (<.001***) | 0.554 (<.001***) | 0.695 (<.001***) | 0.455 (<.001***) | 0.818 (<.001***) | 0.482 (<.001***) |
| Dummy$_{(year)}$ | -0.018 (0.011**) | -0.209 (0.152) | -0.018 (0.023**) | -0.174 (0.181) | -0.017 (0.045**) | -0.091 (0.562) |
| **Dummy$_{(group)}$** | **-0.008 (0.316)** | **0.133 (0.370)** | **-0.009 (0.299)** | **0.245 (0.061*)** | **-0.009 (0.340)** | **0.241 (0.128)** |
| F statistics | 28.24 (<.001***) | 32.08 (<.001***) | 16.43 (<.001***) | 32.30 (<.001***) | 17.48 (<.001***) | 15.48 (<.001***) |
| R-sq. | 0.398 | 0.429 | 0.309 | 0.468 | 0.349 | 0.322 |
| Adjusted R-sq. | 0.384 | 0.416 | 0.291 | 0.454 | 0.340 | 0.301 |
| N | Leader:66; Control:66 | | Leader:57; Control:57 | | Leader:51; Control:51 | |

*Table 2. The Results of Value Network-based Firms*

| Types of IT  Parameter | Aggregate IT Capabilities | | Externally-focused IT Capabilities | | Internally-focused IT Capabilities | |
|---|---|---|---|---|---|---|
| | ROA | Q Ratio | ROA | Q Ratio | ROA | Q Ratio |
| Intercept | 0.050 (0.021**) | 0.212 (0.353) | -0.012 (0.696) | 0.236 (0.166) | 0.042 (0.001***) | 0.139 (0.623) |
| ROA$_{(t-1)}$ or Q$_{(t-1)}$ | 0.328 (<.001***) | 0.943 (<.001***) | 0.407 (0.003) | 0.897 (<.001***) | 0.161 (0.002***) | 0.984 (<.001***) |
| Dummy$_{(year)}$ | -0.015 (0.565) | -0.340 (0.126) | 0.042 (0.260) | -0.209 (0.320) | -0.012 (0.402) | -0.193 (0.391) |
| **Dummy$_{(group)}$** | **-0.012 (0.604)** | **0.241 (0.275)** | **0.010 (0.776)** | **0.232 (0.282)** | **0.008 (0.556)** | **0.230 (0.311)** |
| F statistics | 11.23 (<.001***) | 78.15 (<.001***) | 3.41 (0.023**) | 68.63 (<.001***) | 4.57 (0.007***) | 22.53 (<.001***) |
| R-sq. | 0.403 | 0.824 | 0.150 | 0.780 | 0.238 | 0.606 |
| Adjusted R-sq. | 0.367 | 0.814 | 0.106 | 0.769 | 0.186 | 0.579 |
| N | Leader:27; Control:27 | | Leader:31; Control:31 | | Leader:24; Control:24 | |

*Table 3. The Results of Value Shop-based Firms*

Finally, according to our classification rules, we separately test the three sup-groups (i.e., network, chain, and shop) in the three scenarios of IT capabilities. The results are shown in Table 2 to 4. In Table 2 (value network), the results show that only the externally-focused IT capabilities-enabled firm value (p-value=0.061) is significant and other five tests are not significant. The findings confirm our H3a. That is, as for value network-based firms, their best development strategy of IT capabilities is to strengthen their externally-focused IT capabilities to create superior firm value. In Table 3 (value shop), unfortunately, the results are all insignificant in the six tests. Then, H3b cannot be supported. In Table 4 (value chain), the results show that only the externally-focused IT capabilities-enabled firm performance (p-value=0.042) and the aggregate IT capabilities-enabled firm performance (p-value=0.002) are significant. In addition, the aggregate IT capabilities-enabled firm value is also nearly significant (p-value=0.117). Overall, H3c may be supported and can regarded as another best strategy. That is, as for value chain-based firms, they should balance to develop overall IT capabilities, which include externally-focused and internally-focused IT capabilities.

| Types of IT<br>Parameter | Aggregate<br>IT Capabilities | | Externally-focused<br>IT Capabilities | | Internally-focused<br>IT Capabilities | |
|---|---|---|---|---|---|---|
| | ROA | Q Ratio | ROA | Q Ratio | ROA | Q Ratio |
| Intercept | 0.035<br>(<.001***) | 0.417<br>(0.007) | 0.043<br>(<.001***) | 0.373<br>(0.009***) | 0.023<br>(0.011**) | -0.023<br>(0.855) |
| ROA$_{(t-1)}$ or Q$_{(t-1)}$ | 0.418<br>(<.001***) | 0.780<br>(<.001***) | 0.239<br>(<.001***) | 0.801<br>(<.001***) | 0.754<br>(<.001***) | 1.121<br>(<.001***) |
| Dummy$_{(year)}$ | -0.016<br>(0.080*) | -0.352<br>(0.035**) | -0.012<br>(0.149) | -0.237<br>(0.110) | -0.023<br>(0.013**) | -0.354<br>(0.010***) |
| **Dummy$_{(group)}$** | **0.019<br>(0.042**)** | **0.274<br>(0.117)** | **0.026<br>(0.002***)** | **0.190<br>(0.217)** | **-0.004<br>(0.685)** | **0.036<br>(0.803)** |
| F statistics | 12.90<br>(<.001***) | 164.90<br>(<.001***) | 10.00<br>(<.001***) | 188.75<br>(<.001***) | 20.75<br>(<.001***) | 320.10<br>(<.001***) |
| R-sq. | 0.185 | 0.744 | 0.136 | 0.749 | 0.266 | 0.848 |
| Adjusted R-sq. | 0.171 | 0.740 | 0.129 | 0.745 | 0.253 | 0.845 |
| N | Leader:87; Control:87 | | Leader:97; Control:97 | | Leader:88; Control:88 | |

*Table 4. The Results of Value Chain-based Firms*

## 5 Discussion

Although four of total eleven hypotheses are insignificant (H1e, H1f, H2b, and H3b), we can find out that they are all relevant to the internally-focused IT capabilities. Therefore, the definition of the internally-focused IT capabilities may be further reviewed. First, because they are more related to the quality of IT infrastructure, Bhatt and Grover (2005) believed these qualities are hard to be differentiated and may not be relevant to a firm's advantages. Similar, Chae et al. (2014) also hold the same viewpoint and noted that the highly standardized and homogeneous IT resources may be the reason that firms are difficult to get differentiation performance from IT. Mithas et al.'s (2013) empirical study also shows that there is no evidence for the effect of IT on profitability through operating cost reduction. This implies that cost-related IT implementations are hard to enable greater social complexity, path dependence, and organizational learning and higher barriers to erosion and then realize their value (Mithas et al. 2012). Next, ROA is a firm-level performance indicator. However, the effects of the internally-focused IT capabilities may be more relevant to process-level performance. Therefore, process performance may be more suitable for capturing their benefits (Dehning et al. 2007; Melville et al. 2004).

### 5.1 Research Contributions

This study first attempts to propose a conceptual model to link IT capabilities, industry types, and value implications. Although these constructs are usually examined by previous IT value-related literatures, they almost focus on only one or two constructs. Almost no studies considered and theorized them at the same time in the past. However, this is an important issue when the value creation processes of IT resources and IT capabilities should be further clarified (Schryen 2013).

Next, the development of IT capabilities is usually regarded as the "general purpose" (Luo et al. 2014) nature of IT. However, we believe that this argument may be further revised to respond to the differences between various industries. Although industry-level factors in particular are increasingly important for IS activities (Chiasson and Davidson 2005; Crowston and Myers 2004), the mechanisms of IT value in various industries are still unclear. For example, many studies only shows that IT will have more effects on some industries but not on another industries (e.g., Anderson et al. 2006; Im et al. 2001). They failed to further advice what can these industries, which IT has less effects on, do given each industry all needs IT? This study attempts to answer this important question.

## 5.2 Managerial Implications

Based on our empirical findings, there are at least two implications for managers. First, they must to use the right indicators to measure their investment of IT resources. For example, if they use ROA to measure their IT value, the results will be not expected better than using Q ratio. This finding is also similar to Kohli et al.'s (2012) and Ong and Chen's (2014) conclusions. The worse thing is that IT value will not be always captured if value network-based firms use ROA to measure the effects of the externally-focused IT capabilities. Then, they may possibly make the decision to stop their IT-related investment. Therefore, the linkages between types of IT capabilities and their benefits are the first implications for managers. Next, although H3b is not supported, H3a and H3c are two important IT strategies for value network-based firms and value chain-based firms, separately. That is, the empirical findings advice that value network-based firms should develop externally-focused IT capabilities to create its firm value and value chain-based firms should develop aggregate IT capabilities to improve its firm performance and create its firm value. In general, firms are encouraged to develop their unique portfolio of IT capabilities to maximize their IT value according to their value creation logic.

## 5.3 Conclusions and Limitations

This study attempts to combine the three constructs of IT capabilities, industry types, and value implications. We introduce a new classification methodology of industries (i.e., the value creation logics) and derive three different development strategies of IT capabilities for the three value creation logics of industries. Although the empirical findings cannot support the strategy of the value shop-based industries, we believe that other types of IT capabilities (e.g., information management capability (Mithas et al. 2011) or knowledge management capability) may be suitable for them. Overall, our results provide new insights about how firms could realize their IT value by allocating the portfolios of IT capabilities. Nevertheless, some issues may be improved in the future.

First, although we only can collect the IW data form 2000 to 2001, we believe the data is still valuable and can bring us some insights. Further, the data can be compared to Chae et al.'s (2014) data. Chae et al. (2014) collected IW data form 2001 to 2004. They both are the similar period. However, Chae et al. (2014) only show that IT capability-enabled performance is not significant. We advance their work to illustrate when IT capability-enabled performance is significant and when it is not. Second, this is the first time that the classifications of value creation logics are used to conduct an empirical study. Whether the samples can appropriately represent all the value creation logics must be carefully considered. Or, some industries may have two different value creation logics at the same time (only one domination value creation logics is chosen in this study). These issues should be considered rigorously in the future. However, we believe that the idea of linking new classification methodology to IT value is more important and thought-provoking. Third, the data of the independent variable was collected from InformationWeek and it may have some biases because this ranking can be influenced by the market value, firm performance, and other factors. Nevertheless, the secondary data also provide an opportunity to observe the contributions of IT capabilities in various industries. We also attempt to control some important control variables (e.g., firms' scale, industry type). Forth, because only firms with revenues exceeding 1,000 million dollars are considered in this study, the empirical findings do not reflect the situation for small firms. Although these SMEs may use IS more effectively, this issue should also be noted as a limitation. Finally, this study only focus on the externally-focused and internally-focused IT capabilities to build IT strategies for various industries. Maybe there are other types of IT capabilities that also can explain industry-level variances and the development strategy of IT capabilities. If these types can be addressed well, our argument will be examined more rigorously.

## Acknowledgements

This research is financially supported by the Ministry of Science and Technology of the Republic of China (MOST 103 - 2410 - H - 002 - 094).